\documentclass{PoS}
\usepackage{amsmath}

\newcommand{\betaLC}{\beta_{\mathrm{L}}^{~\mathrm{c}}}

\newcommand{\gL}{g_{\mathrm{L}}}

\newcommand{\gLC}{g_{\mathrm{L}}^{~\mathrm{c}}}

\newcommand{\gLR}{g_{\mathrm{L}}^{~\mathrm{ref}}}
\newcommand{\gTC}{g_{\mathrm{T}}^{~\mathrm{c}}}

\newcommand{\LI}{\Lambda_{\mathrm{L}}^{\mathrm{imp}}}
\newcommand{\EI}{\Lambda_{\mathrm{E}}^{\mathrm{imp}}}

\newcommand{\gIRFP}{g_{\mathrm{4l}}^{\mathrm{IRFP}}}
\newcommand{\gSD}{g_{\mathrm{SD}}^{\mathrm{c}}}

\title{Phases of many flavors QCD: lattice results}

\ShortTitle{Phases of many flavors QCD: lattice results}

\author{Albert Deuzeman\\
        Albert Einstein Center for Fundamental Physics - University of Bern, Switzerland \\
        E-mail: \email{deuzeman@itp.unibe.ch}}

\author{\speaker{Maria Paola Lombardo},~~Kohtaroh Miura\\
        INFN-Laboratori Nazionali di Frascati, I-00044, Frascati (RM), Italy \\
        E-mail: \email{lombardo@lnf.infn.it} \email{miura@lnf.infn.it}
}

\author{Tiago Nunes da Silva,~~Elisabetta Pallante\\
Centre for Theoretical Physics, University of Groningen, 9747 AG, Netherlands \\
Email: \email{e.pallante@rug.nl} \email{tiagoj.nunes@gmail.com}}

\abstract{This note is based on our recent results 
on QCD with varying number of flavors of fundamental fermions.
Topics  include unusual, strong dynamics in the preconformal, confining phase, the physics of the conformal window and the role of ab-initio lattice simulations in establishing our current knowledge of the phases of many flavor QCD.}

\FullConference{Xth Quark Confinement and the Hadron Spectrum\\
                 8?12 October 2012\\
                 TUM Campus Garching, Munich, Germany}

\begin{document}

\section{Phases of Strong Interactions from the Lattice}

While strong interactions spontaneously
break chiral symmetry in  ordinary QCD at zero temperature,
 chiral symmetry is realised either at
high temperatures -- in the so called quark-gluon plasma (QGP)
phase -- and at zero temperature for a large number of flavours $N_f > N_f^c$
\cite{SCGT}.
In the latter case,
the theory is expected to become not only chirally but also 
conformally invariant,  due to the emergence of
an infra-red fixed point (IRFP)
for $N_f > N_f^c$ which prevents the coupling from
growing large enough to break chiral symmetry. 
The main phenomena under
scrutiny are non-perturbative :  only lattice analysis affords the possibility
of  ab-initio studies which are indeed  being carried out 
by many groups\cite{Giedt:2012it}. 

In our studies  we have  identified two main themes of interest:
the physics of the near--conformal window, and the observation of
the conformal window. I will discuss them in the light
of the results we have obtained in recent years. 

\section{Near-conformal : continuum and lattice}

In the near-conformal region\cite{Miura:2011mc,Miura:2012zqa} 
we are mostly concerned
with precursory effects of conformality when approaching $N_f^c$ 
from the QCD side. Model studies suggest three possible scenarios:
an essential singularity a la Miransky--Yamawaki \cite{Miransky:1997}
$1/\xi = exp (- \frac {\pi}{2} \epsilon \sqrt{|Nf - N_f^c|})$; 
a power law conventional behaviour \cite{BraunGies}
  $1/\xi = K | N_f - N_f^c|^{-1/\theta}$; 
and a 'jump' into conformality \cite{Antipin:2012sm} . In the two first cases
the approach to conformality is continuous, and one is likely to
observe precursory effects.  The distinction might
not be so clear-cut and combinations of the various behaviours can
be observed as well, as in a weak first order
transition where an apparent power law behaviour ends with
a  small jump  at the true critical point. 

How do we distinguish a QCD-like dynamics from a more exotic one? 
In either cases chiral symmetry is broken. Gauge dynamics, however, can be 
significantly different. 
The coupling might show a so-called walking behaviour: 
at a variance with the ordinary running, which is regulated by 
one unique scale $\Lambda$, a walking behaviour 
is characterized by two different scales : above the UV scale
$\Lambda_{UV}$  the coupling 
runs towards asymptotic freedom, and below the IR scale $\Lambda_{IR}$
it runs towards confinement, being nearly constant (walking)
in between. 

In short, we can observe walking either by assessing the existence
of two different scales $\Lambda_{UV}$ and $\Lambda_{IR}$  and  by 
observing a
pre-critical behaviour when approaching a critical number of flavor
$N_f^c$.
Note that $1/\xi$ in the above expressions denotes 
any physical quantity with a mass dimension, which includes the
critical temperature $T_c(N_f)$. 
This observation paves the way of a study of
conformality based on the analysis of a thermal system. 

\section{Towards conformality : continuum analysis from lattice results}

\begin{figure}
\begin{center}
\includegraphics[width=6.6cm,height=4.0cm]{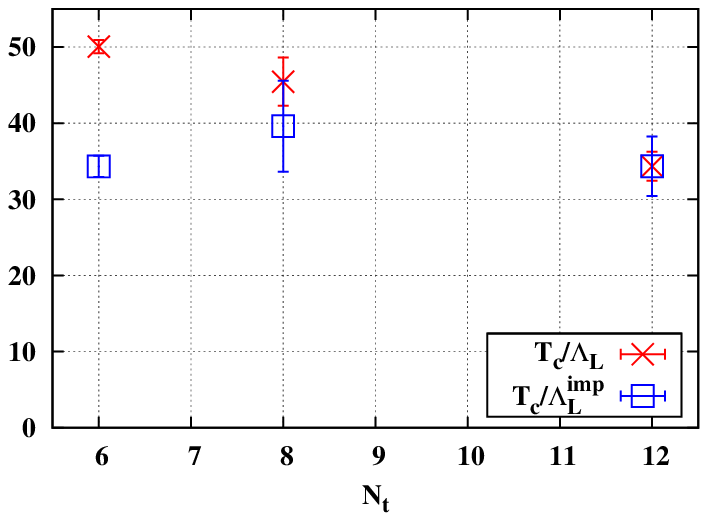}
\includegraphics[width=6.6cm,height=4.0cm]{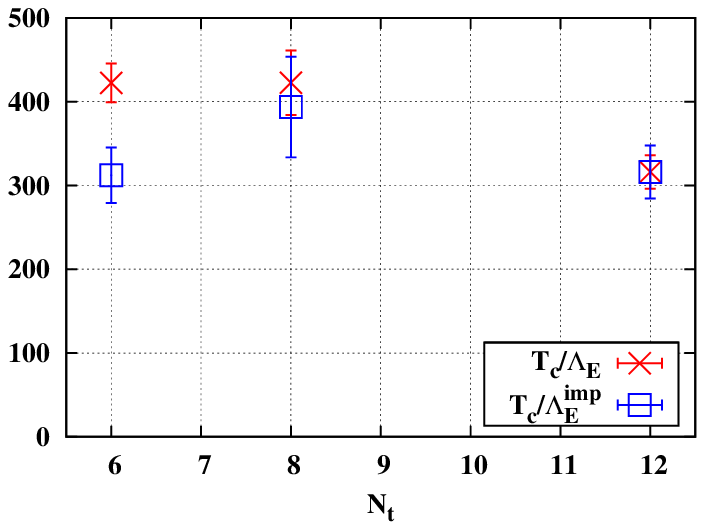}
\caption{Scaling at $N_f = 8$
from the $N_t$ dependence of the normalised
critical temperature.
Left: The bare lattice scheme results.
The red symbol $\times$ shows $T_c/\Lambda_{\mathrm{L}}$,
and the blue $\Box$ symbols represent $T_c/\LI$.
Right: The E-scheme results.
The red symbol $\times$ shows $T_c/\Lambda_{\mathrm{L}}$,
and the blue $\Box$ symbols represent $T_c/\EI$}
\label{Fig:TcL_Nf8}
\end{center}
\end{figure}

A first natural way to highlight a pseudocritical behaviour approaching
conformality is based on the analysis of the $N_f$ dependence of the
critical temperature \cite{Miura:2011mc,Miura:2012zqa}. 
On a lattice the critical temperature is given by  
\begin{equation}
T_c\equiv \frac{1}{a(\betaLC)\cdot N_t}\ .\label{eq:Tc}
\end{equation}
which becomes independent on $N_t$ close to the continuum
limit $a \to 0$.
One  way to convert to physical units 
relies on  the normalised critical temperature
$T_c/\Lambda_{\mathrm{L/E}}$ 
where $\Lambda_{\mathrm{L}}$ ($\Lambda_{\mathrm{E}}$)
represents the lattice (E-scheme) Lambda-parameter
defined in the two-loop perturbation theory
with or without a renormalisation group inspired
improvement. The results for $T_c/\Lambda_{\mathrm{L/E}}$ for the $N_f=8$ 
theory \cite{Deuzeman:2008sc,Miura:2012zqa} 
are shown in Fig. \ref{Fig:TcL_Nf8}, and the full set of results
is collected in Table 1.

\begin{table*}
\caption{
Summary of
$T_c/\Lambda_\mathrm{L}$ and
$T_c/\LI$ for various $(N_f,N_t)$.
The first (second) line at fixed $(N_f,N_t)$
shows the value of $T_c/\Lambda_\mathrm{L}$ ($T_c/\LI$)}\label{tab:TcL}
\begin{center}
\begin{tabular}{c|cccc|cc}
\hline\hline
$N_f\backslash N_t$ &
$4$&
$6$&
$8$&
$12$&
$$&
$$\\
\hline
$0$ &
$18.11\pm 0.65$&
$18.21\pm 0.91$&
$16.56\pm 0.71$&
$$&
$$&
$$\\
\quad &
$16.29\pm 0.75$&
$17.81\pm 1.02$&
$16.56\pm 0.78$&
$$&
$$&
$$\\
\hline
$4$&
$21.99\pm 1.04$&
$19.98\pm 0.95$&
$17.12\pm 2.43$& 
$-$&
$-$&
$-$\\
\quad &
$16.56\pm 1.44$&
$18.67\pm 1.38$&
$17.12\pm 3.41$&
$$&
$$&
$$\\
\hline
$6$ &
$25.41\pm 1.43$&
$25.33\pm 1.43$&
$22.94\pm 1.29$&
$22.30\pm 2.52$&
$$&
$$\\
\quad &
$21.66\pm 1.64$&
$23.87\pm 1.58$&
$22.21\pm 1.40$&
$22.30\pm 2.66$&
$$&
$$\\
\hline
$8$ &
$-$&
$50.05\pm 0.87$&
$47.06\pm 3.28$&
$34.34\pm 1.91$&
$$&
$$\\
\quad &
$-$&
$34.32\pm 1.40$&
$42.67\pm 6.33$&
$34.34\pm 3.90$&
&\\
\hline\hline
\end{tabular}
\end{center}
\end{table*}
From Table 1 we can read  the results for $T_c/\Lambda$ as a function
of $N_f$ in different schemes
($\Lambda = \Lambda_{\mathrm{L}}$ or $\Lambda_{\mathrm{E}}$), which 
consistently show an increase with $N_f$.
This indicates that 
$\Lambda_{\mathrm{L/E}}$  vanishes faster than $T_c$
upon approaching the critical number of flavour. Within the various
uncertainties discussed here, this can be taken as a qualitative
indication of  scale separation close to the critical
number of flavors. 

\begin{figure}
\begin{center}
\includegraphics[width=7.5cm,height=4.6cm]{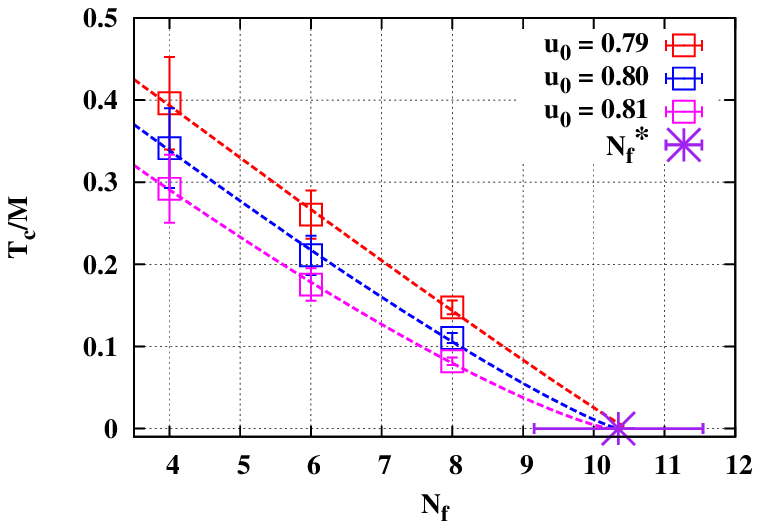}
\includegraphics[width=7.5cm,height=4.6cm]{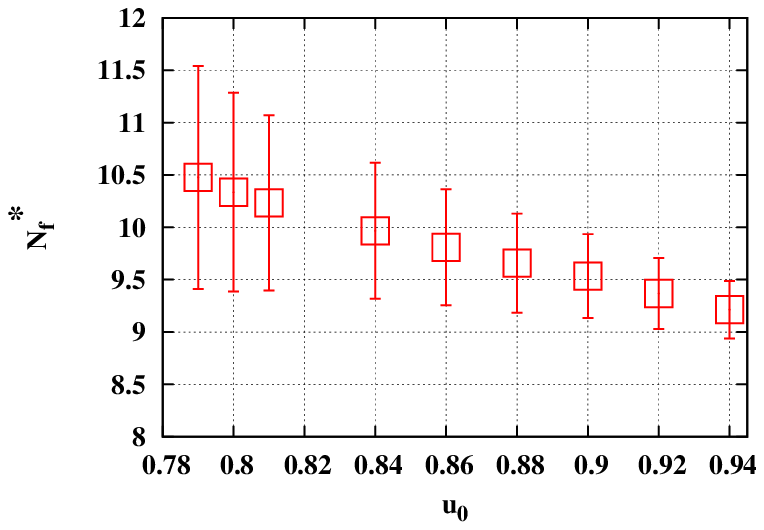}
\caption{
Left:~
The $N_f$ dependence of $T_c/M$ where
$M$ is determined to be a UV scale corresponding to
$u_0=0.79$ (red box),
$0.80$ (blue $\bigcirc$), and
$0.81$ (magenta triangle).
Right: The $u_0$ dependence of $N_f^c$.
The three data in the left side
are determined within the condition $M(\gLR) \lesssim a^{-1}(\gLC)$,
while for the others $M(\gLR)$ exceeds the lattice cutoff.
This more robust procedure confirms our early results, and should
be ultimately confirmed by use a rigorous lattice scale setting
which is in progress \cite{future}.}
\label{Fig:TcM}
\end{center}
\end{figure}

We can now further investigate the vanishing of
the critical temperature. To this end, we have to
face an apparent puzzle :  $T_c/\Lambda$ increases
as a function of $N_f$! This however , as mentioned,
can be understood in terms of scale separation with $\Lambda$
vanishing faster than $T_c$. To see the vanishing of $T_c$ 
at $N_f^c$ we need  to replace $\Lambda$ with a UV scale. 
To do this, we devised a 'baby-version' of the scale setting
procedure in the potential scheme. In that scheme. 
one fixes a value for the renormalised coupling $\bar{g}$
and  $\bar{g}^2\propto r_X^2F(r_X)$ 
sets a scale $r_X^{-1}$. We used our plaquette values to
define a coupling at the scale of the lattice spacing,
and we set a common UV scale for diffeent
theories by imposing a constant value for the
coupling (or, equivalently, for the plaquette).
In short, we use our $u_0 = \langle P \rangle ^{\frac{1}{4}}$  
to define $\bar{g}$,  
and $u_0 = X$ is regarded as the analog of
the potential scheme scale setting.
For this procedure to work, the coupling should be weak enough to
be in the UV region, but also large enough to
avoid major lattice artifacts. We have checked -- the interested
reader is referred to Ref. \cite{Miura:2012zqa} for details --  that it is
possible to meet these requirements and define consistently
an UV coupling over a rather large set of possible choices
of $X$. The results are shown in Fig. \ref{Fig:TcM}

\begin{figure}
\begin{center}
\includegraphics[width=9.5cm]{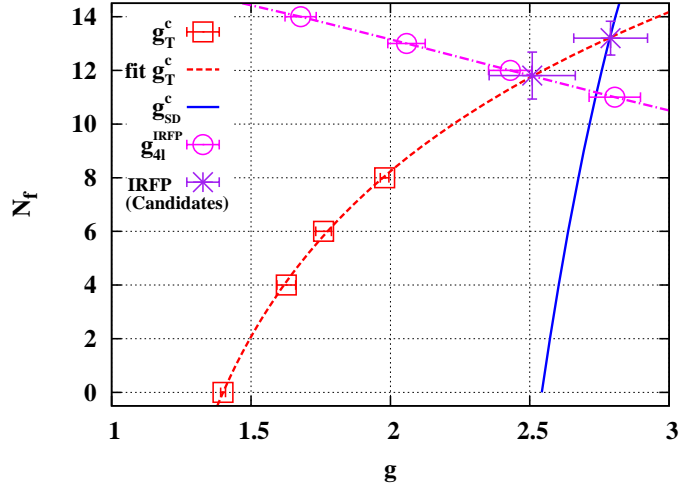}
\caption{
The thermal critical coupling (red $\Box$)
and the fit for them
(dashed red line)
and the values of the zero temperature couplings in the conformal
phase from different estimates, see text for details.
At the critical
number of flavour the thermal critical coupling
should equal the critical
coupling associated with the IRFP.}
\label{Fig:gTC}
\end{center}
\end{figure}

An alternative analysis stems from  a discussion presented in 
Ref. \cite{Liao:2012tw}.  Since the critical temperature is
zero in the conformal phase,
the thermal critical coupling $\gTC$
should equal a zero temperature critical coupling $g^c$
when $N_f = N_f^c$. One possibility 
is to use the Schwinger-Dyson estimate for $g_c$ \cite{Appelquist:1998rb}.
In this case,
the lower edge of the conformal window $N_f^c$
is defined by the condition $\gTC(N_f^c) = \gSD(N_f^c)$ 
We then estimate the intersection of $\gTC$ and $\gSD$ --
hence the onset of the conformal window
as well as the IRFP coupling at $N_f^c$ -- 
at $(g^c,N_f^c) = (2.79,13.2)\pm (0.13,0.6)$.
One second possibility is to 
match $\gTC(N_f^c)$ and  
 the coupling at IRFP
($g^{\mathrm{IRFP}}$) \cite{Ryttov:2012nt}.
We can then locate the intersection of $\gTC$ and $\gIRFP$
and obtain $(g^c,N_f^c) = (2.51,11.8)\pm (0.15,0.9)$.
In Fig.~\ref{Fig:gTC}, we show $g^{\mathrm{IRFP}}$ and $\gSD$ 
alongside with the numerical results for $\gTC$,
as well as the estimates for the IRFP.

\section{Towards conformality : Lattice analysis}

The emergence of the conformal
window can also read-off directly from the lattice data themselves.
Consider the phase diagram 
in the space spanned by the bare coupling $\gL$
and the number of flavor $N_f$, and
the (pseudo)critical thermal lines which connect
the lattice (pseudo)critical couplings for a  fixed $N_t$ . 
Based on the properties of the step scaling function
in the vicinity of a IRFP~\cite{Hasenfratz:2011xn},
it is easy to convince ourselves  
that the critical number of flavor $N_f^c$  can be identified
with the crossing point the pseudocritical thermal
lines obtained for various $N_t$'s

\begin{figure}
\begin{center}
\includegraphics[width=9.5cm]{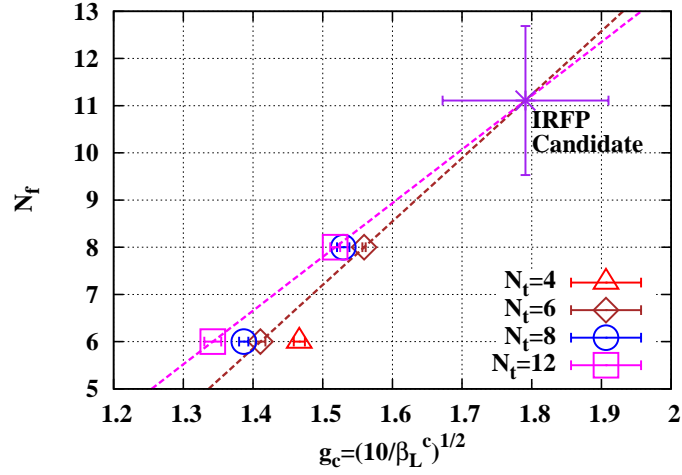}
\caption{(Pseudo) critical values of the lattice coupling
$\gLC=\sqrt{10/\betaLC}$ for theories with $N_f=0,~4,~6,~8$ 
and for several values of $N_t$
in the Miransky-Yamawaki phase diagram.
We have picked up $\gLC$ at $N_f = 6$ and $8$,
and considered  ``constant $N_t$'' lines
with $N_t = 6,\ 12$.
{If the system is still described by
one parameter beta-function in this range of coupling,
the IRFP could be located at the intersection of the
fixed $N_t$ lines }.}
\label{Fig:MY}
\end{center}
\end{figure}

To demonstrate this procedure,
we consider
the pseudocritical lines obtained for $N_t = 6$ and $N_t=12$
as shown in Fig.~\ref{Fig:MY}.
Note their positive slope:
the lattice critical coupling $\gLC$ is
an increasing function of $N_f$.
Interestingly, the slope decreases with increasing $N_t$,
which allows for a crossing point at a larger $N_f$.
Thus, we estimate the intersection at
$(\gLC, N_f^c) = (1.79\pm 0.12,  11.1\pm 1.6)$.
\section{Inside the conformal window: continuum results}

Conformal symmetry implies chiral symmetry. As we are
seeking evidence for conformality in QCD, a natural strategy
is to establish whether the theory realizes chiral symmetry
at zero temperature \cite{Deuzeman:2009mh,Deuzeman:2012pv}. 

A direct observation of chiral symmetry can indeed be
attempted: this means extrapolating the chiral condensate
to the chiral limit. Of course there are obvious numerical
limitations : firstly, the functional form depends on the
realization of chiral symmetry; second, the extrapolated value
is affected by a residual error. So all one can try is
to compare side by side the quality of the extrapolations carried
out with different analytic ans\"atze. While for a small number of
flavors such procedure would unambiguously indicate the breaking
of chiral symmetry, when $N_f$ grows large -- say above $N_f=8$
the results become more ambigous - on which point everyone agrees - 
with some groups favoring chiral symmetry restoration (hence conformality),
and other chiral breaking.

The analysis of the spectrum might offer a more
realiable guidance : it has been noted in the past that
one can devise robust signatures of chiral symmetry based
on the analysis of the spectrum results. 
One first significant spectrum observable is the ratio $m_\pi/m_\rho$, between the mass of the  
lightest pseudoscalar state (pion) $m_\pi$ and the mass of the lightest vector state (rho) $m_\rho$. In QCD at zero temperature, chiral symmetry is spontaneously broken and the pion is the (pseudo)Goldstone boson of the broken symmetry, implying that its mass will behave as $m_\pi\sim \sqrt{m}$.

Within the conformal window chiral symmetry is restored in the continuum limit. 

\begin{figure}
\begin{minipage}[htb]{0.5\linewidth}
\begin{center}
\includegraphics[scale=0.25]{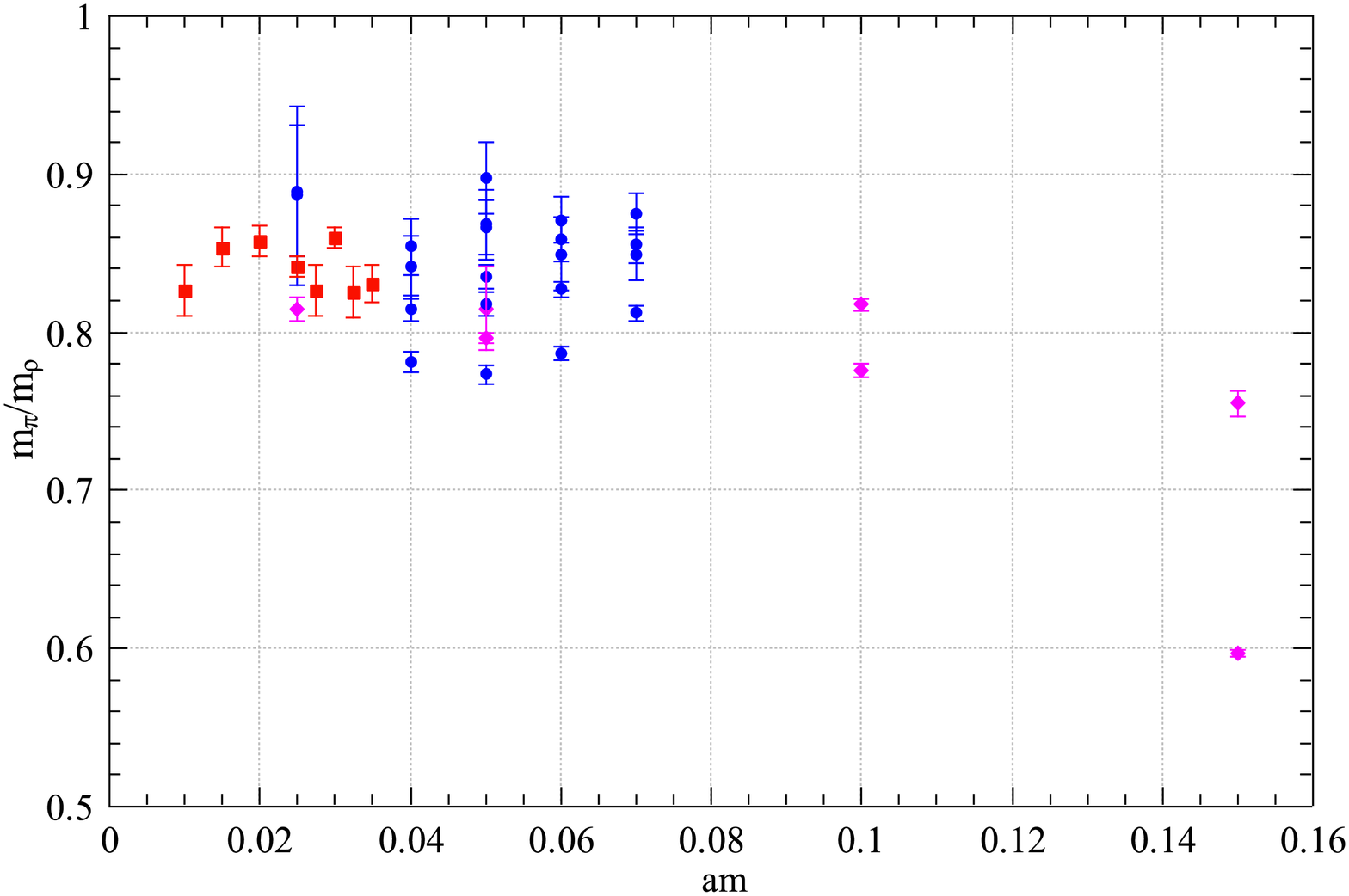}
\caption{Ratio $m_\pi/m_\rho$ as a function of the bare quark mass for all existing data for $N_f=12$, and $N_f=16$: $N_f=12$ data from \cite{Fodor:2011tu} (red squares), $N_f=12$ data from this work and $\beta_L=3.8,3.9,4.0$ (blue circles), $N_f=16$ data  from \cite{Damgaard:1997ut} (magenta diamonds).   }
\label{fig:mpimrhoRatio}
\end{center}
\end{minipage}
\hspace{0.2cm}
\begin{minipage}[htb]{0.5\linewidth}
\begin{center}
\includegraphics[scale=0.25,angle=270]{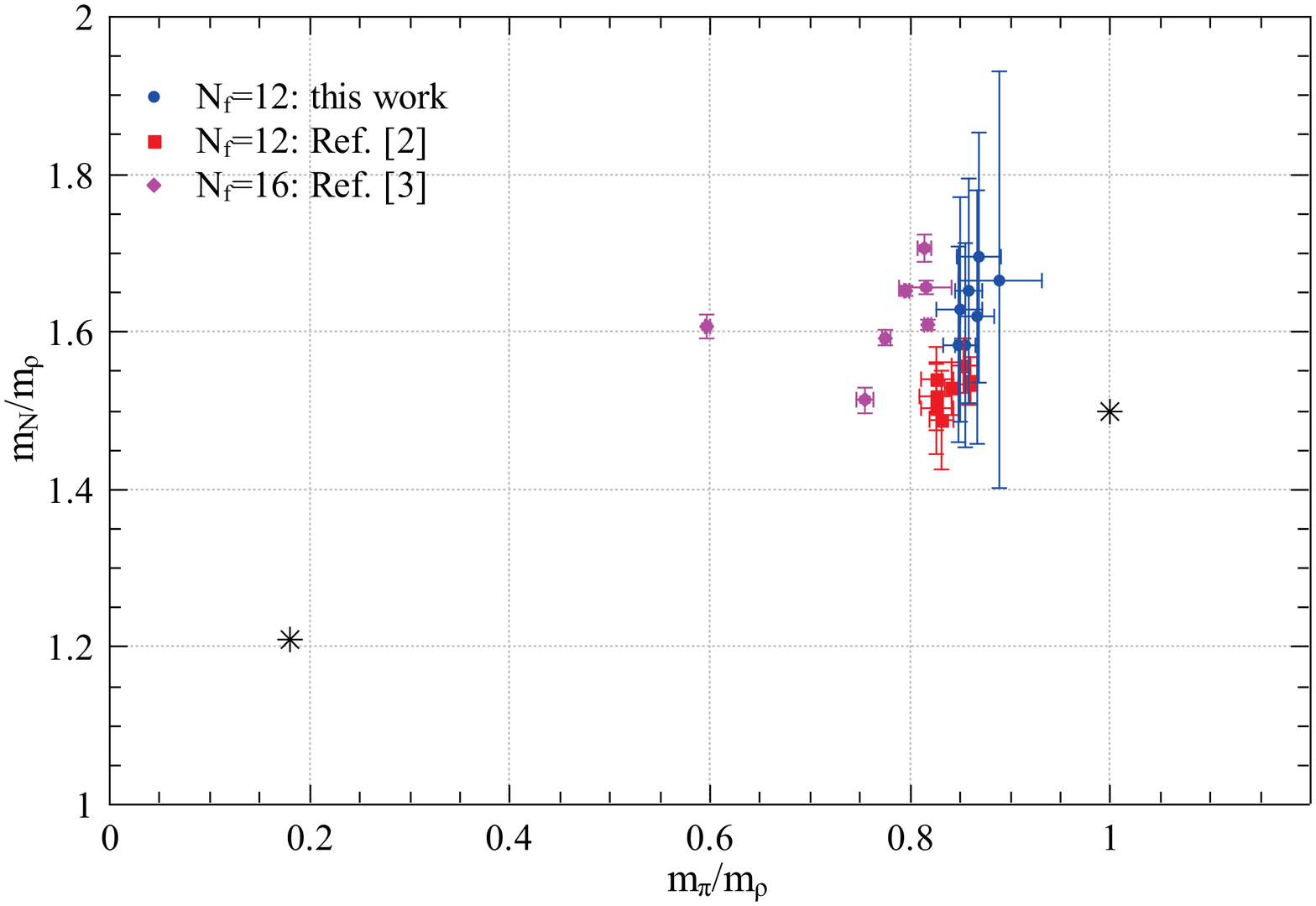}
\caption{Edinburgh plot: $N_f=12$ data from \cite{Fodor:2011tu} (red squares), $N_f=12$ data from this work and $\beta_L=3.8,3.9$ (blue circles), $N_f=16$ data from \cite{Damgaard:1997ut} (magenta diamonds). The QCD physical point (black star, leftmost) and the heavy quark limit (free theory) point (black star, rightmost) are shown.}
\label{fig:Edplot}
\end{center}
\end{minipage}
\end{figure}

At the IRFP and at infinite volume, the quark mass dependence of all hadron masses in the spectrum is governed by conformal symmetry: at leading order in the quark mass expansion all masses follow a power-law with common exponent determined by the anomalous dimension of the fermion mass operator at the IRFP. 
Hence we expect a constant ratio.  Away from the IRFP, for sufficiently light quarks and finite lattice volumes, the universal  power-law dependence receives corrections, due to the fact that the theory is interacting but no longer conformal. The behaviour of the ratio is demonstrated in Fig. \ref{fig:mpimrhoRatio}:
a conformal scenario seems favoured in the range of masses we are
exploring.  Note that the  $m_\pi/m_\rho$ 
ratio should  go to zero in the chiral limit in the broken phase, and
to a constant value if chiral symmetry is restored.

Analgous conclusions can be drawn from the inspection of
the so called Edinburgh plot (\ref {fig:Edplot}). The difference with the
case of ordinary QCD is indeed striking. The modest scattering of the
data points 
could be ascribed to the deviation from a  perfect power law as discussed
above. It would then be of interest to repeat the same plot
for different couplings : at the IRFP it should indeed reduce to a point.
\section{Inside the conformal phase : lattice}

If we were to use a perfect action the conformal phase discussed
above would extend all the way till the infinite coupling limit.
With a naive action instead chiral symmetry appears to be always
broken in the strong coupling limit , at least till $N_f$ is not
too large. An obvious consequence of this is the occurrence of
a strong coupling zero temperature transition -- a bulk transition --  
within the conformal window. The role of improvement in this case 
is really dramatic! A perfect action would destroy a phase transition.
No suprise, of course:
these are strong coupling phenomena taking place away
from the continuum limit, hence extra terms in the actions which are
irrelevant in the continuum might well become relevant.

But then, how would an ordinary improved action (as opposed
to a perfect action) affect the phase transition? The
evidence we have so far is in this case \cite{daSilva:2012wg,Deuzeman:2012ee} 
the bulk transition moves
towards stronger coupling (consistently with the fact that
it will eventually disappear  with a  perfect action) ,
and a second transition develops. Among these two transitions
we have a phase with an unusual realization of chiral symmetry,
observed also in other studies\cite{Cheng:2011ic}.

From the perspective of the analysis of continuum many flavor QCD
these observations are just due to a peculiar form of lattice artifacts.
Bulk transitions are however interesting for several 
reasons including  fundamental 
QFT questions like the existence of an interacting, non--trivial UV 
fixed point in four dimension away from the perturbative domain 
as well as modeling  of condensed matter systems, such as graphene,
and the new phases discussed here might well be of interest in
these contexts. 

\section{Summary}

In brief summary,  we have studied the physics of the
near-conformal window and observed a likely scale separation for
$N_f > 6$. We have developed suitable extrapolation techniques
and estimate in several different ways the critical
number of flavors to be
\begin{align}
N_f^c \sim 
\begin{cases}
11.1\pm 1.6\quad &
(\text{from the vanishing thermal scaling of } \betaLC)\ ,\\
12.5\pm 1.6\quad &
(\text{from the approach of $\gTC$ to $\gSD$
and $\gIRFP$})\ ,\\
10.4\pm 1.2\quad &
(\text{from the vanishing of $T_c/M$
with  $M$ a UV scale})\ .
\end{cases}
\label{eq:Nf_IRFP}
\end{align}

Although these estimates obviously lack precision,  they highlight
in a simple way  the emergence of conformality.  

In the conformal window we have shown how the spectrum analysis
can give information on the realization of chiral symmetry,
and we have discussed features of the strong coupling regime
which might be of interest when modeling a rather a wide class
of phenomena including phase transitions in condensed matter.

There are several directions in which this work can, and
hopefully will be extended: a more rigourours scale setting in the
preconformal region is on the way. The IRFP should be clearly observed,
and we are aiming at doing it working within the analytically tractable
$N_f=16$ model. At a more theoretical level, we hope that this analysis
will help clarifying the behaviour of the anomalus dimension in
the vicinity and away from the IRFP.

More generally, the interplay of the cold conformal window phase with
thermal QCD and the physics of the Quark Gluon Plasma is an interesting,
still largely unexplored, field of research  which we hope to further
pursue in a near future.

\end{document}